%% file: 00_main.tex
\documentclass[10pt,conference]{IEEEtran}
\IEEEoverridecommandlockouts
\usepackage{xcolor}
\usepackage{amsthm,amssymb,multirow,amsmath,amsfonts,cite,mathtools}
\usepackage{todonotes}
\usepackage{bm}
\usepackage[ruled,vlined,algo2e]{algorithm2e}
\usepackage{algorithmic}
\usepackage{algorithm}
\usepackage{subcaption}
\usepackage{multirow}
\usepackage{steinmetz}
\usepackage[normalem]{ulem}
\usepackage{orcidlink}
\usepackage{tikz}
\usepackage{nicefrac}
\usepackage[cmintegrals]{newtxmath}
\usepackage{ulem}
\usepackage[T1]{fontenc}

\usetikzlibrary{arrows}

\newsavebox{\mybox}

\newtheorem{lemma}{Lemma}

\newtheorem{theorem}{Theorem}
\newtheorem{prop}{Proposition}

\newtheorem{cor}[theorem]{Corollary}

\def\BibTeX{{\rm B\kern-.05em{\sc i\kern-.025em b}\kern-.08em
    T\kern-.1667em\lower.7ex\hbox{E}\kern-.125emX}}
\begin{document}

\title{Energy-Optimal Sampling for Edge Computing Feedback Systems: Aperiodic Case}
        
\author{\IEEEauthorblockN{Vishnu Narayanan Moothedath}
\IEEEauthorblockA{\textit{Division of Information Science and Engineering} \\
\textit{KTH Royal Institute of Technology}, Sweden \\
\orcidlink{0000-0002-2739-5060} 0000-0002-2739-5060
}
}

\maketitle

\begin{abstract}
We study the problem of optimal sampling in an edge-based video analytics system (VAS), where sensor samples collected at a terminal device are offloaded to a back-end server that processes them and generates feedback for a user. Sampling the system with the maximum allowed frequency results in the timely detection of relevant events with minimum delay. However, it incurs high energy costs and causes unnecessary usage of network and compute resources via communication and processing of redundant samples. On the other hand, an infrequent sampling result in a higher delay in detecting the relevant event, thus increasing the idle energy usage and degrading the quality of experience in terms of responsiveness of the system. We quantify this sampling frequency trade-off as a weighted function between the number of samples and the responsiveness. We propose an energy-optimal aperiodic sampling policy that improves over the state-of-the-art optimal periodic sampling policy. Numerically, we show the proposed policy provides a consistent improvement of more than 10$\mathbf{\%}$ over the state-of-the-art.
\end{abstract}
\input{01_Intro}
\input{02_sysModel}
\input{04_Aperiodic}
\input{06_ComparisonConclusion}
\section{Acknowledgement}
This research was supported by the Swedish Foundation for Strategic Research (SSF) under the grant ITM17-0246 (ExPECA).
\bibliographystyle{IEEEtran}
\bibliography{refs.bib}
\end{document}

%% file: 01_Intro.tex
\section{Introduction}\label{sec:intro}
Features of the next-generation mobile networks like the releases 15 and 16 of 5G-NR brought with it an increased interest in realising real-time services and applications\cite{alriksson2020critical}. For instance, URLLC (ultra-reliable low latency communication) targets sub-millisecond end-to-end delay demanded in an industrial setting.
Within the class of such delay and latency-sensitive applications, a subgroup of new applications that process snapshots of reality and provide feedback either to devices or humans are receiving an increasing amount of recent research attention.
Some examples of such feedback systems are human-in-the-loop applications such as augmented reality, wearable cognitive assistants (WCA)\cite{TowardsWCA,munoz2020impact}, and ambient safety.
Another example from the domain of cyber-physical systems (CPS) is in the context of automated fault detection, where the acoustic data is processed for vibration analysis to potentially initiate some maintenance, safety or emergency procedures \cite{acoustic}. 
A typical characteristic of these applications is that the feedback quality depends on the timely capture and processing of the state changes via these snapshots, whereas the state changes themselves can be random events. 
Therefore, an efficient sampling of the application is essential in these systems. 
It is even more emphasised by the recent trend of remotely placing most of the processing logic of such feedback systems in edge computing facilities connected with direct wireless links. Such a placement leverages supposedly ubiquitous real-time compute capabilities, however, with an added cost for offloading compute tasks in terms of 
communication delays and energy consumption. 

We investigate systems that employ sampling to monitor a process but only respond to a subset of samples that result in system changes, such as a new augmentation towards a human user in a WCA. These samples are associated with some events of importance, referred to as \textit{essential events}. Other samples do not contain information on such essential events and are ignored.
Following the detection of an essential event from a sample, feedback is generated, and the system transits to the next state where it begins monitoring for an essential next event.
The trade-off that we study relates to the strategy applied to sample the process. Ideally, one aims to have a system that samples the process only once -- immediately after the event completion. However, the a priori information about the event completion required for such a system breaks the causality and makes it infeasible. In any feasible system, the sampling is done with some sampling policies, which only have a statistical idea about the event completion times. 
Any policy that uses more frequent sampling ensures that the crucial event is timely captured.
However, it also results in the capture of insignificant samples of the process, squandering energy, communication bandwidth, and compute cycles.
In this work, we examine approaches that enable the prompt capture of relevant system changes in an edge-based feedback system while also minimising overall energy usage.

Event detection from control theory literature typically looks at event-triggered control where an event occurs when the sensor detect that a reading has crossed a threshold\cite{eventTriggeredControl}. However, these studies are not applicable to our case because they do not rely on any necessary assumptions regarding the amount of the data being communicated, the requirement of feedback for control, or the remote processing and detection of events.
Works like \cite{veeravalli2012quickest,ananthanarayananreal,edgebox,videoSurveillanceSurvey} that contains these assumptions
are mostly based on the quickest detection of the events. Many of them do not take the aspect of energy consumption into consideration, a perspective which is becoming increasingly important. 
Those that consider this aspect mostly come from the video analytics and surveillance domain where multiple strategies to reduce energy usage are discussed. These include optimising sensor topologies \cite{topology}, optimising video coding and transmission techniques \cite{VidCoding}, forcing sensor cooperation \cite{NodeCooperation}, and selective frame transmission or sensor activation\cite{energyefficientobjectdeteciton,smartSleeping}.
In contrast, we reduce the data generated at the sensors by statistically determining the optimum sampling instants, thereby reducing the total amount of data in the communication and processing pipeline. 

A different but well-studied approach to saving energy is offloading the sensor data.
By making wise offloading decisions for the samples, the disadvantage of an increased delay accumulated up on a large number of samples during offloading is somewhat mitigated.
\cite{survey,offloading_MEC1}. This includes binary decisions \cite{JayaLiang_SemiOnlineAlgos,JayaLiang_singleRestart2,offloading_binary1,offloading_binary3}, partial offloading decisions \cite{energyAwareOffloading,offloading_partial1}, and stochastic decisions \cite{offloading_stochastic}. 
While these works focus on the sensor side of the system but not on the total energy usage which is simply shifted to the edge device. However, our work implements a framework for minimising the total energy usage of the system by reducing the number of samples collected, transmitted and processed.

In our previous works, we have extensively studied the efficient capture of essential events in a video analytics system (VAS) and a general cyber-physical system (CPS) using an optimal periodic sampling\cite{OptSampling_iccSage,optSampling_tmc}.
The VAS in our research is motivated by the WCA from \cite{TowardsWCA,munoz2020impact} where a human task progress is monitored continuously by a video stream processed at a remote server for the detection of a predefined task completion. The feedback generated after the task completion is used to assist a human user in continuing with the remaining tasks that together complete a whole process. 
The energy-optimal periodic sampling policies that we proposed provided considerable improvement in energy efficiency over a baseline policy considered. However, an obvious unanswered question that was kept aside for future research in these works was the potential for further improvement by removing the periodicity constraint and looking at the class of more generic aperiodic sampling policies.

In this work, we propose an optimal aperiodic sampling policy that can further reduce energy usage in an edge-based feedback system. To find this policy, we retain a large portion of the system model but remove those parts that mandate the periodicity of the sampling policy under consideration. This forces us to follow completely different mathematical tools and approaches. We use a two-step approach where we first solve for the optimum sampling instants given the time of the first sample, and then find the optimum first sampling instant using an efficient algorithmic approach.
The idea of such an approach is adapted from the checkpointing literature in computing systems\cite{CP_A}. 
In this work, we prioritise Rayleigh distribution in our mathematical formulations. This is because, past works on WCA \cite{TowardsWCA,munoz2020impact} and our own distribution fitting using task completion time dataset from \cite{munoz2020impact} suggest that the task completion times follow Rayleigh distribution.

The key contributions of our work are listed below.
\begin{enumerate}
    \item We    
    propose an energy-optimal aperiodic sampling policy for a general distribution of task completion times.
    \item We prove the convergence of the two-step solution approach for Rayleigh distributed task completion times.
    \item Using simulations, we show that the energy usage under the optimal aperiodic policy is lower by $10$\% compared to that under the optimum periodic sampling policy.
\end{enumerate}
The rest of this paper is organised as follows. In the next section, we discuss the system model. Section~\ref{sec:optSampling} contains the solution and convergence proof followed by the simulation results in Section \ref{sec:compare}.
We conclude in Section \ref{sec:conclude}.

%% file: 02_sysModel.tex
\section{System Model and Problem Statement}\label{sysmodel}
\begin{figure}[t]
\centerline{\includegraphics[width=0.98\linewidth]{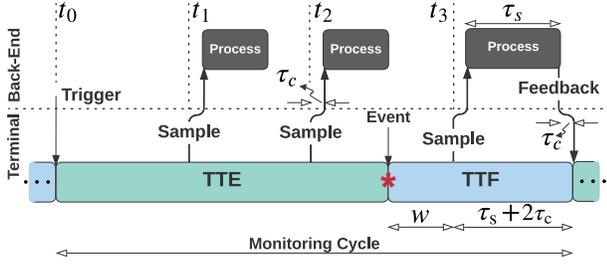}}
\caption{Timing diagram of an arbitrary monitoring cycle.}
\label{fig:model}
\end{figure}

We consider an edge feedback system consisting of a terminal and a back-end server (referred to as simply terminal and server from here on), that together monitor a random process via sampling it. The sensor at the terminal captures and sends the samples to the back-end server for processing. 
For example, in the WCA system studied in \cite{ChenZhuoHu,munoz2020impact}, the user is asked to complete a predetermined set of tasks -- for instance, assembling a set of Lego pieces -- and the essential events correspond to the completion of each task. Each of these tasks takes a random amount of time for completion. An image sensor (which is the terminal or in the terminal) takes images (or video frames) of the user activity and sends them to the back-end for processing via a wireless channel.
Immediately after the essential event, the process (or the human user in the above example) transitions to a temporary state where no more events are expected. The next sample drawn after this transition point -- referred to as a \textit{successful sample} -- will trigger an event detection at the back-end's processor which then provides feedback to the terminal indicating the task completion. The reception of this feedback marks the start of a fresh \textit{monitoring cycle} and the process continues. Only the successful sample results in the generation of feedback, while all other samples are discarded at the back-end. The time taken from the start of a monitoring cycle to the event occurrence is referred to as \textit{time to event} or \textit{TTE}, and the time between this event and the reception of the corresponding feedback at the terminal is referred to as \textit{Time to feedback} or \textit{TTF}. 
In Fig. \ref{fig:model}, we show the timing diagram corresponding to one monitoring cycle of such a system.

Sampling the system to detect an event is controlled by a sampling policy which is a set of sampling instants denoted by $\{t_n,n\!\geq\!1\}$. This includes both the successful sample that triggers the feedback as well as all the discarded samples taken during the TTE. 
The TTE and the total number of samples are denoted by the random variables $\mathcal{T}$ and $\mathcal{S}$, respectively. 
The TTF consists of a random wait time $\mathcal{W}$ between the event occurrence and the immediate next sample, a deterministic processing delay $\tau_{\text{s}}$ of the successful sample, and a two-way communication delay denoted by $2\tau_{\text{c}}$. 
Let a realisation of $\mathcal{W}$ be $w$.
It is important to note that the processing and communication of the successful sample alone contributes to the TTF, while that of the discarded samples occur during the TTE occur within the TTE.
The terminal device enter into an \textit{idle mode} when not performing any transmission or reception, incurring an idle power consumption of $P_0$ (typically much less than $P_{\text{c}}$). 
In this work, we assume for the sake of simplicity that the total power consumption $P_{\text{c}}$ is the same at both the terminal and back-end during transmission and reception of samples or feedback. We also assume that the communication delay $\tau_{\text{c}}$ is the same in both directions, and that the processing delay at the back-end is smaller than the sampling interval. The latter assumption adds simplicity by avoiding duplicate sampling after the event completion. Let $f_X(.)$, $F_X(.)$, and $\bar{F}_X(.)$ denote the PDF, CDF, and CCDF of random variable $X$, respectively.
The notations are summarised in TABLE \ref{tab:notations}. 
\begin{table}[t]
\centering
    \begin{tabular}{|llll|}
    \hline
    $\mathcal{S}$&number of samples&
    $\mathcal{W}$&wait time\\
    
    $\mathcal{T}$& time to event (TTE)&
    $t_n $&$n^{\mathrm{th}}$ sampling instant\\
    
    $\tau_{\text{c}} $&communication delay&
    $\tau_{\text{s}} $&processing delay\\
    
    $P_{\text{c}} $&communication power&
    $P_0$&idle power\\
    
    $\mathcal{E}$&energy penalty&
    ${f}_\mathcal{X}(\cdot)$&PDF $X$\\
    
    ${F}_\mathcal{X}(\cdot)$&CDF $X$&
    $\bar{F}_\mathcal{X}(\cdot)$& CCDF of $\mathcal{X}$\\
    \hline
    \end{tabular}
    \caption{Table of notations.}
    \label{tab:notations}
\end{table}

As discussed in the previous section, an ideal policy samples the system immediately after the occurrence of an event so that there is exactly one sample and the wait $w\!=\!0$. However, such a policy is infeasible given the randomness of the TTE.
Thus, one has to settle with a policy that finds a balance between the expected number of samples $\mathbb{E}[\mathcal{S}]$ and the expected wait $\mathbb{E}[\mathcal{W}]$ to minimise the total energy consumption. Each sample consumes energy in terms of communication and processing, and idle energy is expended during the wait $w$. We quantify this energy usage as a function of the sampling instants $\{t_n\}$ and find a set that minimises the expected energy usage.
Note that, $\mathcal{T}$ is a system property whereas $\mathcal{S}$ and $\mathcal{W}$ -- are derived from $\mathcal{T}$ through the selection of $\{t_n\}$. 
We can compute the energy $\mathrm{E}$ required at the terminal to detect an event as \cite{optSampling_tmc}
\begin{alignat}{2}
    \mathrm{E}&=(\mathcal{S}+1)\tau_cP_c+(\mathcal{T}+\mathcal{W}+\tau_\mathrm{s}+2\tau_\mathrm{c}-(\mathcal{S}+1)\tau_c)P_0\nonumber\\
    &=\mathcal{S}\tau_{\text{c}} (P_{\text{c}} -P_0)+\mathcal{W}P_0+(\mathcal{T}+\tau_{\text{c}} +\tau_{\text{s}} ) P_0+\tau_{\text{c}} P_{\text{c}}\,.\nonumber 
\end{alignat}
Here, the terms except the first two containing the random variables $\mathcal{S}$ or $\mathcal{W}$ have constant expectations for a fixed distribution of $\mathcal{T}$. Thus, these terms are irrelevant to the energy optimisation. Define \textit{energy penalty} $\mathcal{E}\big(\{t_n\}\big)$ or simply $\mathcal{E}$ as the expectation of $\mathrm{E}_{\mathrm{r}}$, the relevant components of energy, where
\begin{alignat}{1}
\mathrm{E}_{\mathrm{r}}&=\mathcal{S}\tau_{\text{c}} (P_{\text{c}} -P_0)+\mathcal{W}P_0.\nonumber\\
\Rightarrow\mathcal{E}&=\mathbb{E}(\mathrm{E}_{\mathrm{r}})=\alpha\mathbb{E}[\mathcal{S}]+\beta\mathbb{E}[\mathcal{W}],\;\label{eq:epsilon_terminal}
\end{alignat}
where $\alpha\!=\!\tau_{\text{c}} (P_{\text{c}}\!-\!P_0)$ and $\beta\!=\!P_0$ are constants. Here, $\alpha\mathbb{E}[\mathcal{S}]$ and $\beta\,\mathbb{E}[\mathcal{W}]$ corresponds to the energy wasted per discarded sample and the additional energy expended for waiting, respectively. In this work, we study the optimisation problem to find the optimum policy $\Pi^*$ such that,
\begin{alignat}{1}
\Pi^*: \{t_n^*\}=\underset{\{t_n^*\}}{\text{arg\,min}}\;\mathcal{E}\big(\{t_n\}\big)\,.
\end{alignat}
In practice, the solution is computed once prior to starting the sampling, for a given distribution of the TTE. This computation can be done as part of the admission control procedures at the back-end
or at the sensor -- if it is capable of it. The proposed solution does not involve any additional signalling overhead because of this one-time a priori computation.
Furthermore, it is interesting to note that, the following mathematical analysis can be applied not only to optimise energy but also to optimise other metrics written in the form \eqref{eq:epsilon_terminal}, simply by adapting the constants $\alpha$ and $\beta$.\color{black}

%% file: 04_Aperiodic.tex
\section{Optimal Sampling}\label{sec:optSampling}
In this section, we find the optimum set of sampling instants $\{t_n^*\}$ for a given TTE distribution. First, we find $\{t_n^*,n\!\geq\!2\}$ recursively for a given $t_1$ and then find $t_1^*$ using an algorithm, an approach inspired from \cite{CP_A}. Next, we demonstrate and prove the convergence of the algorithm. Although most of the following analysis is valid for a general TTE distribution, we give specific focus to the relevant Rayleigh distribution for proofs, wherever necessary. Recall from section \ref{sec:intro} that the relevance of Rayleigh distribution is motivated by previous works on WCA as well as from distribution fitting.

\subsection{Recursive Solution}
Define $t_0=0$ and recall that $\alpha$ and $\beta$ are the penalty weights. If the TTE realises at $\mathcal{T}\!=\!t$ such that $t_n<t\!\leq\! t_{n+1}$, we have, 
\begin{alignat}{1}
    \mathrm{E_r}\big(\{t_n\}\,\vert\,\mathcal{T}=t,t_n<t\leq t_{n+1}\big)=\alpha(n+1)+\beta(t_{n+1}-t).\nonumber\\
   \Rightarrow \mathcal{E}=\mathbb{E}(\mathrm{E}_{\mathrm{r}})=\sum_{n=1}^{\infty}\int_{t_{n-1}}^{t_n}\!\big(\alpha n+\beta(t_{n}-t)\big)f_\mathcal{T}(t)\,\mathrm{d}t\label{ExpEnPen}.
\end{alignat}
It is trivial that the sequence of sampling intervals dictated by the sampling instants should be strictly positive. Furthermore, we observe that $\mathcal{E}$ does not converge when this sequence $t_n\!-\!t_{n-1}$ is increasing in nature\cite{CP_A}. Thus, we restrict the set of sampling instants to a set that satisfies the conditions of
\begin{subequations}\label{cond}
\begin{alignat}{3}
\!\!&(a)\text{ positive sampling intervals}\,&:&\, t_{n+1}\!-\!t_n\!>\!0,\text{ and}\label{cond.1}\\
\!\!&(b)\text{ decreasing sampling intervals}\,&:&\, t_{n+1}\!-\!t_n\!<\!t_{n}\!-\!t_{n-1}.\label{cond.2}
\end{alignat}
\end{subequations}
The sequences of sampling intervals -- or equivalently, sampling instants -- that satisfy these conditions are referred to as \textit{valid sequences}. 
It is interesting to note that the condition \eqref{cond.2} is satisfied for the optimum samples of any general distribution that is a Pólya frequency function of order 2 \cite{theoryOfreliability,CP_A}. One easy check for such distributions is the increasing nature of the hazard function $\nicefrac{f_\mathcal{T}(t)}{\bar{F}_\mathcal{T}(t)}$ which is true for a Rayleigh distribution, thus confirming the existence of a solution.
Now, we differentiate \eqref{ExpEnPen} with respect to $t_n$ and equate them to zero for all $n$. To find the derivative, we use the Leibniz rule for integration, where only the $n^{\text{th}}$ and $(n\!+\!1)^{\text{th}}$ terms of~\eqref{ExpEnPen} produce a non-zero result. 

Thus we have,
\begin{multline*}
    \dfrac{\partial}{\partial t_n}\mathcal{E}\big(\{t_n\}\big)=\,\dfrac{\partial}{\partial t_n}\int_{t_{n-1}}^{t_n}\!\big(\alpha n+\beta(t_{n}-t)\big)f_\mathcal{T}(t)\,\mathrm{d}t\\+\dfrac{\partial}{\partial t_n}\int_{t_{n}}^{t_{n+1}}\!\!\!\big(\alpha(n+1)+\beta(t_{n+1}-t)\big)f_\mathcal{T}(t)\,\mathrm{d}t
\end{multline*}
\begin{alignat*}{1}
=&\;\alpha n f_\mathcal{T}(t_n)+\!\!\int_{t_n}^{t_{n+1}}\!\!\!\!\beta f_\mathcal{T}(t)\,\mathrm{d}t-\big(\alpha(n+1)+\beta(t_{n+1}-t_n)\big)f_\mathcal{T}(t_n)
\end{alignat*}
\begin{alignat*}{1}
=&\,\beta\big(F_\mathcal{T}(t_n)-F_\mathcal{T}(t_{n-1})\big)-f_\mathcal{T}(t_n)\big(\alpha+\beta(t_{n+1}-t_n)\big).
\end{alignat*}
Equating the derivative to zero gives,
\begin{alignat}{1}
t_{n+1}=t_n+\dfrac{F_\mathcal{T}(t_n)-F_\mathcal{T}(t_{n-1})}{f_\mathcal{T}(t_n)}-\dfrac{\alpha}{\beta},\quad\forall n\geq1.\label{recursive}
\end{alignat}
The solution for a Rayleigh distributed TTE can be obtained by substituting the corresponding CDF and PDF in \eqref{recursive}. That~is,
\begin{alignat}{1}
t_{n+1}=t_n+\dfrac{\sigma^2}{t_n}\Big(\mathrm{exp}\big({\tfrac{t_n^2-t_{n-1}^2}{2\sigma^2}}\big)-1\Big)-\dfrac{\alpha}{\beta},\quad\forall n\geq1.\label{recursive_rayleigh}
\end{alignat}
In general, this condition is not sufficient for optimality, but only necessary. 
However, for a given value of $t_1$, \eqref{recursive_rayleigh} provides a unique set of $\{t_n,\,n\!\geq\!2\}$ and hence this necessary condition is sufficient here for determining the optimum sampling instants for a fixed $t_1$. Thus, this recursion reduces the dimension of the search space of the optimum sampling instants from infinity to one and we just have to search for $t_1^*$ -- the optimum $t_1$.

\subsection{Optimum $t_1$}
\begin{figure}[b]
\centerline{\includegraphics[width=\linewidth]{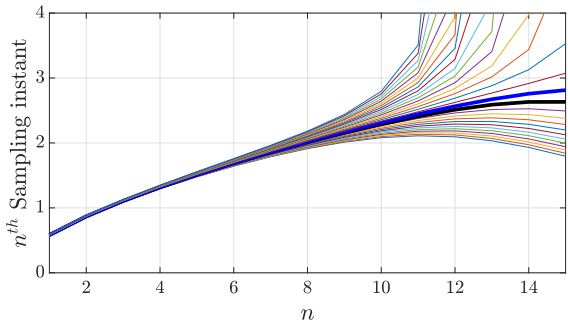}}
\caption{Evolution of $\{t_n\}$ with $n$ generated by the recursion \eqref{recursive_rayleigh} using different values of $t_1$. Two sequences valid upto $n\!=\!15$ ($t_1\!=\!582\,$ms and $t_1\!=\!582.5\,$ms) are highlighted with bold lines.
}
\label{fig:tkVsk}
\end{figure}
 Let $t_n(t_1)$ be the $n^{\mathrm{th}}$ sampling instant and $\{t_n(t_1)\}$ be the set of all sampling instants generated using \eqref{recursive_rayleigh} by an arbitrary~$t_1$. Also, let $\mathcal{E}(t_1)\!:=\!\mathcal{E}\big(\{t_n(t_1)\}\big))$.
 To find $t_1^*$, we start with a discussion on the nature of these sequences of sampling instants given by \eqref{recursive_rayleigh}.
Before the analytical discussion, we first illustrate their typical behaviour using Fig. \ref{fig:tkVsk} where we plot a few sequences $\{t_n(t_1)\}$ versus $n$ for a Rayleigh distributed TTE with $\mu\!\!=\!\!1\,$s. We consider the first $15$ samples and use $\nicefrac{\beta}{\alpha}\!\!=\!\!21$, the reason for which will be explained later in section \ref{sec:compare}. Adjacent lines show the sequences obtained with consecutive $t_1$ in the chosen list of $t_1$ from $577\,$ms to $590\,$ms that differ by $0.5\,$ms. We can see that the sequences with smaller $t_1$ violate \eqref{cond.1} as n goes to $15$ and the graph starts to decrease continuously. Similarly, the sequences with larger $t_1$ close to $0.59\,$s eventually violate \eqref{cond.2} and the graph goes up towards infinity. 
 In this illustration, only two sequences with $t_1\!=\!582\,$ms and $t_1\!=\!582.5\,$ms (highlighted with bold lines) are valid up to $n\!=\!15$, even though more sequences are valid for a lesser $n$.
 It can be inferred that a sequence $\{t_n\}$ generated using \eqref{recursive_rayleigh} by any $t_1$ may not a valid sequence and that the validity may be very sensitive to small changes in $t_1$. In the following, we establish a few characteristics of these sequences.
\begin{lemma}\label{lemma10}
Consider a Rayleigh distributed TTE with parameter $\sigma$. If $t_1^{({1})}$ and $t_1^{({2})}$ are two finite starting sampling instants such that $t_1^{({1})}<t_1^{({2})}$, then we have $t_n(t_1^{({1})})\leq t_n(t_1^{({2})}),\,\forall n\geq2$. 
\end{lemma}
\begin{proof}
The partial derivative of $t_{n+1}$ can be obtained from \eqref{recursive_rayleigh}.
\begin{alignat}{1}
\dfrac{\partial t_{n+1}}{\partial t_{n}}=\;&1+\dfrac{\sigma^2}{t_n^2}\Big(\dfrac{t_n^2}{\sigma^2}\,\mathrm{exp}\big({\tfrac{t_n^2-t_{n-1}^2}{2\sigma^2}}\big)-\mathrm{exp}\big({\tfrac{t_n^2-t_{n-1}^2}{2\sigma^2}}\big)+1\Big)\nonumber\\
=\;&1+\mathrm{exp}\big({\tfrac{t_n^2-t_{n-1}^2}{2\sigma^2}}\big)-\dfrac{\sigma^2}{t_n^2}\Big(\mathrm{exp}\big({\tfrac{t_n^2-t_{n-1}^2}{2\sigma^2}}\big)-1\Big).\nonumber
\end{alignat}
Assume that the partial derivative $\tfrac{\partial t_{n+1}}{\partial t_{n}}$ is non-positive. That~is,
\begin{alignat}{1}
\bigg(\dfrac{t_n^2}{\sigma^2}\bigg)\dfrac{\mathrm{exp}\big({\tfrac{t_n^2-t_{n-1}^2}{2\sigma^2}}\big)+1}{\mathrm{exp}\big({\tfrac{t_n^2-t_{n-1}^2}{2\sigma^2}}\big)-1}&\leq1\nonumber\\
\Rightarrow\;\bigg(\dfrac{t_n^2-t_{n-1}^2}{2\sigma^2}\bigg)\dfrac{\mathrm{exp}\big({\tfrac{t_n^2-t_{n-1}^2}{2\sigma^2}}\big)+1}{\mathrm{exp}\big({\tfrac{t_n^2-t_{n-1}^2}{2\sigma^2}}\big)-1}&<1\label{lemmaEq1}\\
\text{Let }x=\dfrac{t_n^2-t_{n-1}^2}{2\sigma^2}\quad\Rightarrow\dfrac{x\,(e^x+1)}{(e^x-1)}&<1.
\end{alignat}
However, we can easily see that $\tfrac{x\,(e^x+1)}{(e^x-1)}\!>\!2,\,\forall x,$ thus forming a contradiction and invalidating the initial assumption. Hence,
\begin{subequations}\label{posPartialDer}
\begin{alignat}{1}
\dfrac{\partial t_{n+1}}{\partial t_{n}}>0,\,\forall n\geq1.\label{posPartialDer_a}\\
\Rightarrow \dfrac{\partial t_{n+1}}{\partial t_{1}}=\prod_{i=1}^n\dfrac{\partial t_{i+1}}{\partial t_{i}}>0,\,\forall n\geq1.\label{posPartialDer_b}
\end{alignat}
\end{subequations}
The proof can be easily completed using \eqref{posPartialDer_b}.
\end{proof}
We claim using Lemma \ref{lemma10} that, if a sequence with a particular $t_1$ violates \eqref{cond.1} and starts to decrease in value, so does any other sequence with a smaller value of $t_1$. 
Similarly, if a sequence with a particular $t_1$ violates \eqref{cond.2} and starts to increase towards infinity, so does any other sequence with a larger value of $t_1$. 
This claim can be supported using the following arguments.
Let $\{t_n(\hat{t_1})\}$ violates \eqref{cond.2}. That is, $t_n(\hat{t_1})\!\rightarrow\!\infty$ for some large n. Now assume a $t_1\!>\!\hat{t_1}$. According to Lemma~\ref{lemma10}, this implies that $t_n(t_1)\!\geq\!t_n(\hat{t_1}),\,\forall n\!\geq\!2$. As a result,  $t_n({t_1})\!\rightarrow\!\infty$ for some large n thus implying a violation of \eqref{cond.2}. This same argument can be extended for those sequences that violate \eqref{cond.1}.
In other words, it is the smaller values of $t_1$ that generate a sequence potentially violating \eqref{cond.1}, and it is the larger values of $t_1$ that generate a sequence potentially violating \eqref{cond.2}. We write this formally in the below corollaries.
\begin{cor}\label{cor2}
If $\{t_n(\check{t_1})\}$ violates \eqref{cond.1}, so does $\{t_n(t_1)\},\forall t_1\!\!<\!\check{t_1}.$ Similarly, if $\{t_n(\hat{t_1})\}$ violates \eqref{cond.2}, so does $\{t_n(t_1)\},\,\forall t_1\!>\!\hat{t_1}$.
\end{cor}
\begin{cor}\label{cor1}
If $\{t_n(\check{t_1})\}$ violates \eqref{cond.1} and $\{t_n(\hat{t_1})\}$ violates \eqref{cond.2}, then $t_n(\check{t_1})\leq t_n(\hat{t_1}),\,\forall n\geq1$. 
\end{cor}
\begin{algorithm}[b!]
\SetAlgoLined
 Initialise the range of optimising variable, $\check{t_1}$ and $\hat{t_1}$;\\
 Initialise stopping criterion $t_{\bar{n}}$.\\
$n\gets1\,;\quad$
$t_0\gets0\,;\quad$
$t_1\gets(\check{t_1}+\hat{t_1})/2\,;$\\
\While{$t_n\leq t_{\bar{n}}$}{\algorithmiccomment{\%\ Bisection iteration to find optimum $t_1$}\\
$n\gets1\,;\quad$
$t_1\gets(\check{t_1}+\hat{t_1})/2\,;$\\
\While{$1$}{
\algorithmiccomment{\%\ Recursion to find $\{t_n^*\,n\!\geq\!2\}$ for the current $t_1$}\\
$t_{n+1}=t_n+\tfrac{\sigma^2}{t_n}\big(\mathrm{exp}({\tfrac{t_n^2-t_{n-1}^2}{2\sigma^2}})-1\big)-\tfrac{\alpha}{\beta}\,;$\\
\eIf{$t_{n+1}-t_{n}<0$}
    {$\check{t_1}\gets t_1\,;$\\\textbf{break}}
    {
        \If{$t_{n+1}-t_{n}>t_{n}-t_{n-1}$}
            {$\hat{t_1}\gets t_1\,;$\\
            {\textbf{break}}}
    }
$n\gets n+1\,;$
}
}
$t_1^*\gets t_1\,;\quad$
$\mathcal{E}^*\gets \mathcal{E}(t_1)\,;$
\caption{Algorithm to find optimum sampling instants.}\label{Algo}
\end{algorithm}

We use Algorithm~\ref{Algo} which is inspired by the bisection algorithm to compute $t_1^*$. We start the algorithm by assigning the lower and upper limits to two arbitrary $\check{t_1}$ and $\hat{t_1}$, as in the corollaries. The limits are then repeatedly updated whenever the sequence generated by the bisection variable becomes invalid; based on whether \eqref{cond.1} is violated, or \eqref{cond.2}. We will now discuss the optimality of $t_1^*$ obtained using the algorithm.

\begin{prop}
The result of the algorithm $\mathcal{E}^*
$ is arbitrarily close to the infimum achievable energy penalty, given the bounded differentiability of $\mathcal{E}$ with respect to $t_1$.
\end{prop}
\begin{proof}
Define an  invalid $t_1$ as a $t_1$ that generates an invalid sequence using \eqref{recursive_rayleigh}. It is clear from the corollaries that any $t_1$ smaller than a $t_1$ violating \eqref{cond.1} or any $t_1$ larger than a $t_1$ violating \eqref{cond.2} is invalid. Hence, an invalid $t_1$ cannot exist between two valid $t_1$. In other words, the set $\{t_1\!:\! \{t_n(t_1)\}\text{ is valid}\}$ containing $t_1^*$ forms a non-disjoint interval. Thus the initial search space $[t_n(\check{t_1}),t_n(\hat{t_1})]$ of the Algorithm \ref{Algo} contains $t_1^*$ within it.
As a result, the bisection-inspired algorithm exponentially converges to $t_1^*$, and a bounded differentiablity of the energy penalty suggests that $\underset{t_1\rightarrow t_1^*}{\lim}\mathcal{E}(t_1)\!=\!\mathcal{E}(t_1^*)$.
\end{proof}

Checking the bounded differentiability of $\mathcal{E}$ analytically is hard due to the recursion involved. However, we have verified it using simulations, when the TTE is Rayleigh distributed. Until now, we have discussed about infinite length sequences of sampling instants. However, the definition of a finite-length valid sequence is tied with an $\bar{n}$ below which the validity is maintained and is necessary for practical purposes. An invalid sequence can be made valid by considering only a finite part of it, with a length less than $\bar{n}$. For instance, the two valid sequences in Fig. \ref{fig:tkVsk} has $\bar{n}\!>\!15$. We use this threshold $n\!=\!\bar{n}$ to terminate the algorithm such that the probability of the TTE taking a value above $t_{\bar{n}}$ is as low as one wants.

To further take care of a potential TTE realisation greater than $t_{\bar{n}}$ (however small it may be), we can adapt the policy by allowing one final sample at a very large $t$, after the termination of the algorithm.
For this purpose, we choose a small enough probability value $\epsilon$ such that the realisations of the TTE above $\bar{F}^{-1}(\epsilon)>>t_{\bar{n}}$ can be neglected. 
Note that, $t_{\bar{n}}$ and $\epsilon$ are fixed a priori by the user irrespective of the initial value $t_1$ or the algorithm, whereas $\bar{n}$ is obtained by the algorithm for a given $t_1$ and $t_{\bar{n}}$.
Define $\hat{\mathcal{E}}$ as the error in the expected penalty incurred as a result of stopping the algorithm at $t_{\bar{n}}$ and not considering a potential TTE realisation of  $\mathcal{T}\in (t_{\bar{n}},\bar{F}^{-1}(\epsilon))$ for optimisation. 
That is,
\begin{alignat*}{1}
\hat{\mathcal{E}}&=\mathbb{P}\big(t_{\bar{n}}<\mathcal{T}\leq\bar{F}^{-1}(\epsilon)\big)\cdot\big(\alpha+\beta \hat{w}\big),
\intertext{where $\hat{w}$ is the wait when $t_{\bar{n}}<\mathcal{T}\leq\bar{F}^{-1}(\epsilon)$. Note that}
\hat{w}&\leq\bar{F}^{-1}(\epsilon)-t_{\bar{n}}\\
\Rightarrow\hat{\mathcal{E}}&\leq\big(\bar{F}(t_{\bar{n}})-\epsilon\big)\big(\alpha+\beta (\bar{F}^{-1}(\epsilon)-t_{\bar{n}})\big).
\end{alignat*}
For instance, a decent $t_{\bar{n}}\!>\!6\mu$ and a very small $\epsilon\!=\!10^{-22}$ for a Rayleigh distributed $\mathcal{T}$ give us $\hat{\mathcal{E}}\leq6(\alpha+2\beta\mu)\times10^{-13}$. 
This is negligible compared to the typical penalty values. Note~that, $\hat{\mathcal{E}}$ depends on $\epsilon$ and $t_{\bar{n}}$ but not on the algorithm or $\bar{n}$. 
One can repeat the algorithm until either $n = \bar{n}$ or until $t_1^{({2})}\!-t_1^{({1})}$ comes below the computation precision of the system.
Recall the illustration in Fig. \ref{fig:tkVsk} where the valid $t_{15}\!\approx\!2.8\,$s with $\bar{F}_\mathcal{T}(2.8)\!\approx\!0.002$. For $\epsilon\!=\!10^{-22}$, this results in $\hat{\mathcal{E}}\!\leq\!0.0037$.

%% file: 06_ComparisonConclusion.tex
\section{Performance comparison}\label{sec:compare}
In this section, we illustrate the working and performance of the proposed optimal sampling policy $\Pi^*$ in minimising $\mathcal{E}$. We compare the resultant $\mathcal{E}$ with that obtained using a baseline policy $\Pi_\mathrm{b}$ and the state-of-the-art optimal periodic policy $\Pi_\mathrm{p}$ -- all applied on a practically relevant VAS mentioned in section~\ref{sysmodel}. 
The characterisation of the VAS is motivated by the {Google Glass}\cite{googleGlass} and from the WCA experiments in \cite{munoz2020impact}.
These experiments use a frame size around $300\,$kB (that is, a resolution of $640\times 480$) and observe a mean task time of $4.846\,$s.
Google Glass use an 802.11ax transmitter which provides a data rate of $400\,$Mbps which results in a $5.85\,$ms communication delay for each of the $300\,$kB frames. Note that with the terminal located in the proximity of the edge, this contribution of propagation delay is negligible.
Furthermore, the Google Glass consumes a power of 334mW and 2960mW during \textit{active/screen-off} and \textit{video chat}, respectively. 
We take these power figures as the idle power $P_0$ and the communication power $P_\mathrm{c}$ for our simulations, respectively.
These characterisations give us an $\nicefrac{\beta}{\alpha}$ ratio of $21.7$.
For the policy $\Pi_\mathrm{b}$, we choose a sampling interval of $83.3\,$ms which is also motivated by the mean sampling interval of the WCA system in \cite{munoz2020impact}. 
Note that, these are the same characterisation that we used in our previous work \cite{optSampling_tmc} and we reuse them here for consistency.

\begin{figure}[t]
\centerline{\includegraphics[width=1\linewidth]{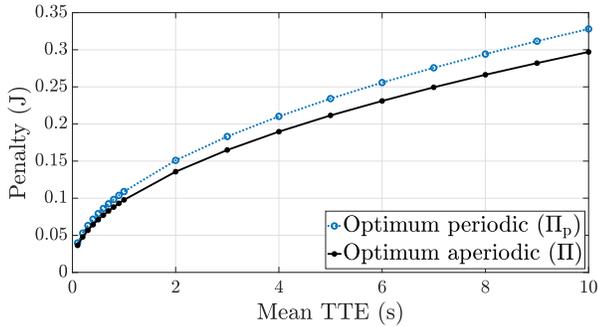}}
\caption{Energy penalty obtained on a VAS by the optimal policy compared with that of an optimum periodic policy for different mean values of the Rayleigh distributed TTE.}
\label{fig:muVsPen}
\end{figure}
\begin{figure}[t]
\centerline{\includegraphics[width=1\linewidth]{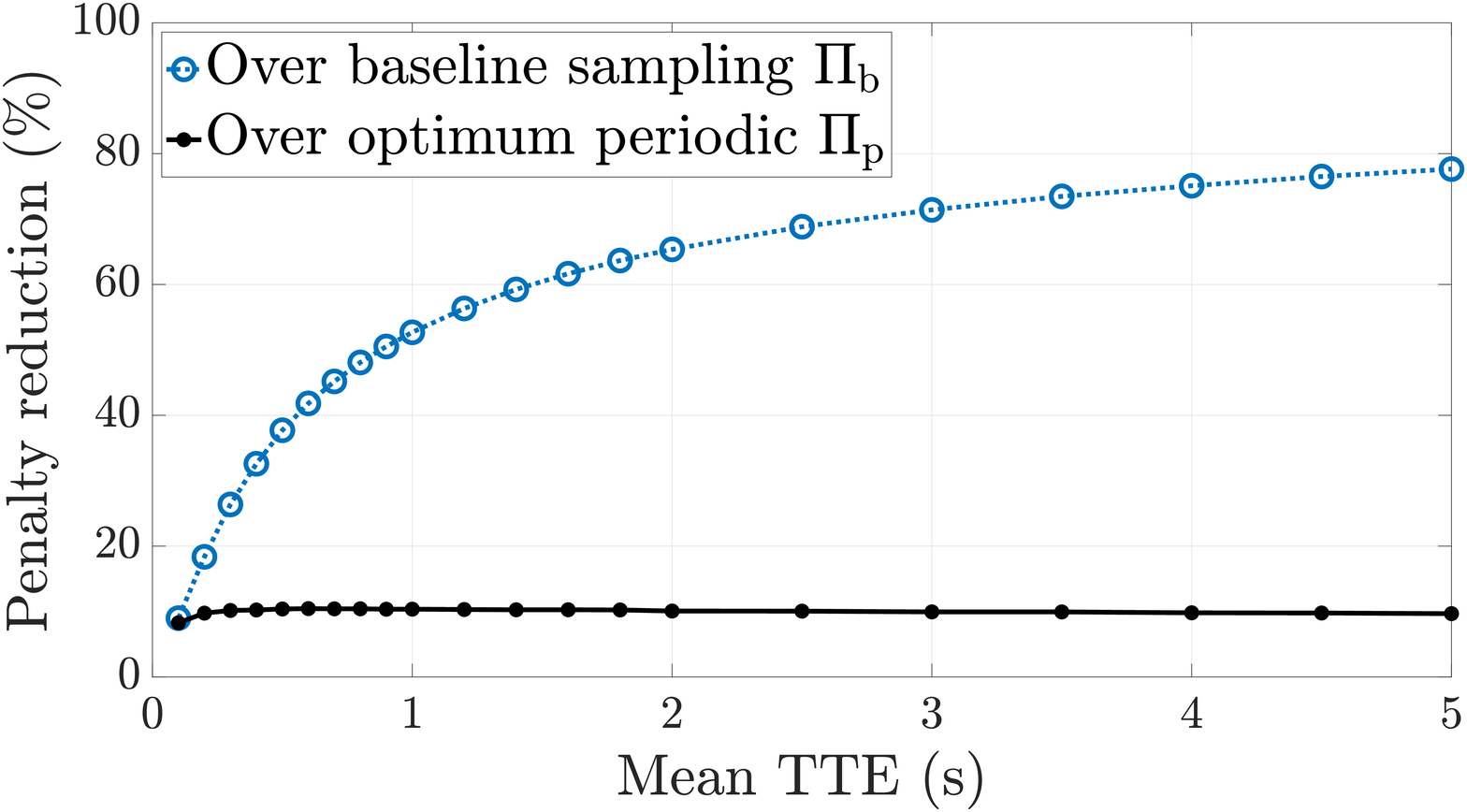}}
\caption{Percentage penalty reduction achieved on a VAS by the proposed policy over the baseline policy and the optimum periodic policy for different mean values of the Rayleigh TTE.}
\label{fig:penaltyreduction2}
\end{figure}
In Fig. \ref{fig:muVsPen}, we compare $\mathcal{E}$ obtained with $\Pi^*$ and $\Pi_\mathrm{p}$ by plotting it against the mean of the Rayleigh distributed TTE. We can see that the proposed policy $\Pi^*$ is consistently performing better than $\Pi_\mathrm{p}$. We did not include the baseline policy $\Pi_{\mathrm{b}}$ in this illustration because, with the large energy improvement already gained with $\Pi_{\mathrm{p}}$ over $\Pi_{\mathrm{b}}$, the improvement achieved on top of that by $\Pi^*$ would have been less apparent. Nevertheless, the additional energy reduction achieved by the proposed policy cannot be undermined. For instance, at $\mu\!=\!5\,$s we see a $9.8\%$ energy penalty reduction attained by $\Pi^*$ over $\Pi_\mathrm{p}$.
To illustrate the increased energy efficiency, in Fig.~\ref{fig:penaltyreduction2} we plot the penalty reduction attained by $\Pi^*$ over $\Pi_\mathrm{b}$ and $\Pi_\mathrm{p}$. The improvement in energy efficiency achieved by $\Pi^*$
continuously increases with $\mu$ over $\Pi_{\mathrm{b}}$, whereas over the optimal periodic policy it stabilises at around $10\%$.

\begin{figure}[t]
\centerline{\includegraphics[width=1\linewidth]{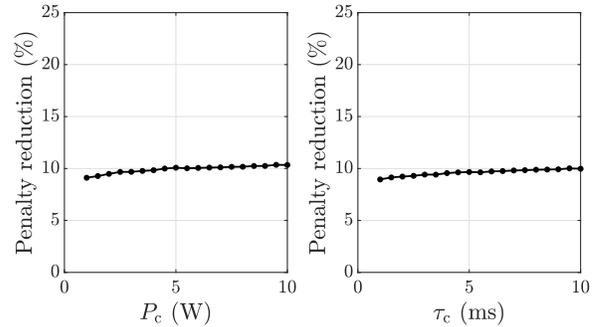}}
\caption{Percentage penalty reduction achieved on a VAS by the proposed policy over the optimum periodic policy for different values of communication delay $\tau_{\mathrm{c}}$ and power $P_{\mathrm{c}}$.}
\label{fig:percVsTcPc_1}
\end{figure}

We have observed that for various mean values, the percentage decrease in energy penalty achieved by using $\Pi^*$ over $\Pi_\mathrm{p}$ is stable at around $10\%$, irrespective of the ratio $\nicefrac{\beta}{\alpha}$. In other words, the proposed policy outperforms the state-of-the-art by a constant amount irrespective of the communication power $P_{\mathrm{c}}$ and delay $\tau_{\mathrm{c}}$ of the application, which is the only parameters apart from the idle power that affects the optimisation. We show this in Fig. \ref{fig:percVsTcPc_1}by plotting the percentage penalty reduction versus the $\tau_{\mathrm{c}}$ and $P_{\mathrm{c}}$ for VAS with a Rayleigh distributed TTE of mean $4.84\,$s. 
We can see the constant $10\%$ improvement discussed above.

\section{Conclusion}\label{sec:conclude}
We considered an edge-based video analytics system (VAS) that captures essential events via sampling.
We proposed an energy-optimal aperiodic sampling policy using a two-step iterative approach. The first step analytically finds the optimum sampling instants for a given time of the first sample and the second step finds the optimum first sampling instant. 
We proved the convergence of the two-step approach and illustrated the consistent performance improvement of the proposed policy over a baseline policy and the state-of-the-art optimal periodic policy. 
